\documentstyle[12pt,aasms,psfig]{article}
\def\etal{{\it et\thinspace al.}\ }

\def\lmlm{{$\lambda\lambda$}\ }

\def\ang{{$\AA$}\ }
\def\mcg6{{MCG--6-30-15}\ }

\def\bl{{$\bullet$}\ }
\def\ka{{\rm K$_{\alpha}$}\ }
\def\chan{{\it Chandra}\ }
\def\xmm{{\it XMM-Newton}\ }
\def\tov{{$\sim 17.5 \AA$}\ }
\def\dfe{{$\frac{df}{d\epsilon}$}\ }
\def\fbar{{$\bar{f}_r$}\ }

\input epsf

\newcommand{\be}{\begin{equation}}
\newcommand{\ee}{\end{equation}}

\begin{document}
\let\typeset\relax

\title{X-RAY RESONANCE OPACITY OF OXYGEN AND IRON IN AGN MCG--6-30-15}

\author{Anil K. Pradhan, Guo-Xin Chen, Franck Delahaye, Sultana N.
Nahar, and Justin Oelgoetz}
\affil{Department of Astronomy, The Ohio State University, Columbus, OH
43210}

\begin{abstract}

 Theoretical differential oscillator strengths related to 
monochromatic opacity due to K- and L-shell absorption from
oxygen and iron ions are directly compared with the \chan and \xmm
spectra of Seyfert~1 galaxy \mcg6. We compute the highly resolved
continuum with resonances due to O~I~-~VI and Fe~XVI.
It is found that the KLn ($n \rightarrow \infty$) resonance series 
limits in O~VI,
and the lowest LMM resonance
in Fe~XVI, lie at the prominent \tov break in the observed spectra. 
We also calculate and identify, for the first time, the observed gap 
in spectral flux at 22-23 \ang in the \chan spectra
due to K $\rightarrow$ L resonant absorption features from {\it all} 
O-ions O~I~-~VI,
potentially leading to abundance determination of several or all ionization 
stages; similar signature gaps may be predicted due to other elements.
The precise atomic parameters are computed
in the relativistic close coupling approximation using the Breit-Pauli
R-matrix method. The new X-ray opacities might possibly distinguish between 
models of a dusty warm absorber, and/or gravitational redshift and
broadening due to a massive black hole in \mcg6.

\end{abstract}


\keywords{X-Rays : general --- Ultraviolet : general --- atomic
processes --- line: formation, identification --- radiation mechanisms: 
thermal}

\section{INTRODUCTION}

 The Seyfert 1 galaxy \mcg6 has been under extensive study with new and
continuing observations from \chan and \xmm. 
The soft X-ray region 
exhibits a predominantly non-thermal continuum with `breaks' or
`spectral turnovers'
with anomalous widths and depths larger than
expected due to lines or photoabsorption `edges' 
(Branduardi-Raymont \etal 2000, Lee \etal 2001,
Sako \etal 2001). The primary interest in \mcg6 stems from eariler
{\it ASCA} observations of the hard X-ray spectrum that 
show the prominent 6.4 keV
iron \ka line with a large redward component, interpreted as gravitational
broadening of emission from the inner region of the accretion disc around
a massive black hole (Tanaka \etal 1995).
Similarly, the positions and widths of the soft X-ray
features were interpreted as gravitational broadening and redshift of
recombination emission of H- and He-like Oxygen, and other elements,
by Branduardi-Raymont \etal (2000) using the \xmm data. However, Lee \etal (2001)
obtained higher resolution \chan data with the HETG and explained the spectrum
in terms of a `dusty warm absorber' model, with Rydberg series of lines
from highly charged oxygen and iron ions, without
requiring relativistic emission. The HETG observations of \mcg6 show a
prominent break or spectral turnover at \tov, at the edges corresponding
to line opacity due to higher members of the Rydberg series 1s$^2$ 
$\longrightarrow$ 1snp of O~VII, and the onset of the Fe-L edges. Their
model also includes Fe~XVII and Fe~XVIII lines in the shorter wavelength 
region $\lambda < 17.5$  \ang. At 22.05 \ang Lee \etal identify
resonant photoabsorption 
by a KLL resonance in O~VI predicted earlier (Pradhan 2000), and a
tentative identification of the O~I resonance at $\sim 23$ \ang. These
observations appeared to corroborate the Lee \etal warm absorber model. 
However, whereas the Lee \etal model includes line opacity from O and Fe 
ions, it does not consider resonance opacity (with the exceptions noted above).

Recently
Sako \etal (2002) have reported additional \xmm RGS observations and
analysis, reconfirming the conclusions of Branduardi-Raymont \etal (2000) as to
gravitational broadening in \mcg6. The \xmm RGS data samples the longer
wavelength region beyond that of the \chan HETG, where Sako \etal find
considerable differences between the observed flux and 
the dusty warm absorber model proposed by Lee \etal 
They also find O~IV and O~V features, in addition to the O~VI feature 
found earlier. Although the Sako \etal model considers opacity 
sources from lines and resonances of the identified species (including
iron oxides), they note in particular that there is insufficient opacity of
O and Fe at the \tov break, and that the Lee \etal
model overestimates the flux compared to observations. Hence they
conclude that the shape and magnitude of the observed features are
inconsistent with photoelectric absorption, and required
Fe and O abundances, and reaffirm the model with relativistically
broadened Ly$\alpha$ lines of ionized C,N, and O in irradiated
accretion discs of \mcg6, as well as another Seyfert 1 galaxy Mrk 766. 

 The two particular aims of this paper are: (i) to demonstrate
resonance opacity from O and Fe ions as a significant contributor 
at \tov (in addition to line opacity), 
and (ii) to identify inner-shell resonant \ka absorption features from all 
O-ions, O~I~-~O~VI, that manifest themselves as the oxygen signature gap
at 22-23 \ang in \mcg6. The general aim is to illustrate the level of detail
and precision required for astrophysical models in order to model \chan
and \xmm sources. Atomic parameters for line, resonant, 
and continnum absorption are computed using ab intio methods with 
relativistic fine structure. The calculated oscillator strengths and
photoionization cross sections are related directly to the monochromatic
opacity spectrum which, in turn, determines the emitted flux and
absorption features.

\section{THEORY AND COMPUTATIONS}

 Resonance opacity complements line and continuum opacity. The
monochromatic absorption coefficient may be defined in terms the
quantity {\it differential oscillator strength} \dfe (e.g. Pradhan 2000, Nahar
\etal 2001) which subsumes these 
forms of atomic opacity, including autoionizing resonances. 
 We express the line (bound-bound) and the resonant $\oplus$ non-resonant
(bound-free) absorption coefficient in terms of \dfe

\be
a^{b-b} (\epsilon) = N_i f = N_i (\frac{\pi e^2}{mc}) (\frac{2z^2}{\nu^3}) 
\frac{df}{d\epsilon}; \ \ \ \ \
a^{b-f} (\epsilon) = N_i \sigma_{PI}(\epsilon)= (4\pi^2\alpha a_o^2) 
\frac{df}{d\epsilon},
\ee

 and where N$_i$ is the ion density, $z$ is the ion charge,
and $\nu$ is the effective quantum number. The differential oscillator
strength \dfe = $(\nu^3/2z^2) \ f_{line}$ for line opacity, and 
\dfe = $(1/4 \pi^2 \alpha a_0^2) \ \sigma_{PI}$, where $f_{line}$ is the
oscillator strength, and $\sigma_{PI}$ is the photoionization cross
section. The \dfe thus enables equal
representation in magnitude of the b-b and b-f absorption coefficients,
and is the basic atomic quantity that determines monochromatic 
opacity. We compute the \dfe for all ions under consideration from 
the oscillator strengths and photoionization cross sections obtained
from elaborate and extensive relativistic close coupling 
calculations using the Breit-Pauli R-matrix (BPRM) method (Scott and
Taylor 1982, Hummer \etal 1993, Zhang \etal 1999).

 All atomic calculations reported in this work are new, and have been
carried out using configuration interaction eigenfunction 
expansions for the seven core O-ions, O~II~-~O~VIII, and for Fe-ions
Fe~XVI, XVII, and XVIII. In addition to the bound-free results reported
herein, line $f$-values are also calculated for O~VI and O~VII, and
Fe~XVI and Fe~XVII, including transitions up to $n$ = 10 levels. 
While the voluminous
atomic data calculations are still in progress, and will be reported
separately, we present the results pertaining to resonances
in the \ka complexes of O~I~-~VI, and L-shell complex of Fe~XVI.

\section{RESULTS}

 In order to focus on the atomic physics of the opacity in the spectral
region of the \tov turnover in the \mcg6 spectra, and the identification of
resonant absorption from \ka complexes of O-ions and the LMM complex of
Fe~XVI, we consider the \dfe for the relevant ions as
obtained from ab initio calculations.
In this report we eschew any modeling or fitting. While the relative 
magnitude of opacities depends
on the densities, temperatures, ionization fractions,
abundances etc. in a model, the actual spectral features --- in energies and
shapes --- should be basically the same as the detailed structure in \dfe.

 Fig. 1 shows the \ka resonant complexes of
oxygen, with inner-shell excitation in all O-ions O~I~-~OVI.
Photoabsorption cross sections are from the ground level, and the dominant
components and the peak positions of resonances in each ion are shown.
In fact each complex has several components; their positions may or
may not coincide depending on the exact energies of the contributing
angular and spin symmetries. For example, O~I and O~II
show one single peak because the contributing final J$\pi$ symmetries all lie 
at the same energy ($\pi$ refers to the parity of the electronic state).
On the other hand resonances begin to spread out with ion charge
in O~III and O~IV, which also present more complicated resonance
structures. The ground level of O~III, $1s^22s^2p^2 (^3P_0)$,
photoionizes into the J = (1)$^o$ continuum of $1s2s^22p^3$ with the 
three components shown. O~V has only one final J$\pi (^1P^o)$ since the
initial level is $1s^22s^2 \ (^1S_0)$. The KLL O~VI is a twin component
system discussed in detail in earlier works (Pradhan 2000, Nahar \etal
2001; the latter work includes Li-like C~IV, O~VI, and Fe~XXIV).
However, the O~VI resonance at 21.87 \ang is about an order of
magnitude weaker than the stronger one at 22.05 \ang; the latter was
identified in the \mcg6 spectrum by Lee \etal (2001; see Fig. 2
bottom panel). Most of these results for O-ions have been obtained for the first
time (prior works on O~I and O~II resonances are discussed in Paerels
\etal 2001, and below). The calculations also include
the \ka resonant oscillator
strengths \fbar = $\int_{\Delta E_r} (df/d\epsilon) d\epsilon$ values for all
O-ions (to be presented later), that might potentially determine
column densities for all ionization stages of oxygen.

 Fig. 2(a) presents the \dfe for O and Fe ions. The O~VI and the Fe~XVI
results are shown including the background, whereas the O~I~-~O~V
resonances are shown only around the peak values that would be manifest
in the observed \mcg6 spectra (bottom panel). The observed absorption spectrum
corresponds to the inverse of the opacity. Therefore in Fig. 2(b) we
plot (\dfe)$^{-1}$, analogous to the $1/\kappa(h\nu)$ term in the
integrand of the Rosseland harmonic mean for stellar opacities (e.g.
Seaton \etal 1994), where $\kappa_{\nu}$ is the monochromatic opacity.
The calculated opacity spectrum matches rather closely the observed
spectra, and the large resonant opacity due to inner-shell
excitation of six O-ions, O~I~-~O~VI, results in the gap or
window in the 22-23 \ang region clearly discernible in the \chan 
\mcg6 spectrum by Lee \etal (2001). 
As observed, the O~I resonance brackets this region
on the longer wavelength side at 23.5 \ang, and the stronger component
of the O~VI KLL at 22.05 \ang on the shorter wavelength side.

 The spectral turnover or break at \tov in Fig. 2 has been discussed by 
Sako \etal
(2002) with respect to both \chan and \xmm observations. Their warm
absorber model
flux is overpredicted compared to observations due to `lack of O~VII and
Fe L opacity' (Sako \etal, Fig. 11). Here we find
that the highest O~VI KL$n$ resonance series limit, as $n \rightarrow \infty$
lies at 17.42 \ang. The resonances, towards the longer wavelength side
$\sim$ 17.5 - 17.8 \ang, converge to several $n = 2$ thresholds $1s2s 
(^3S_1,^1S_0)$ and $1s2p (^3P^o_{0,1,2},^1P^o_1)$. Resonances in O~VI,
up to $n = 50$, have been fully resolved at about 70,000 energies with
an effective quantum number mesh that enables equal resolution for each
$n$-complex. The downward jump in the
theoretical absorpton spectrum in (\dfe)$^{-1}$  at \tov (Fig. 2b) corresponds 
directly to the similar jump in the observed flux. 

 There are actually two discrete jumps as seen
in Fig. 3. The enhancement is mainly due to the considerable oscillator
strengths in the $1s \rightarrow np$
inner-shell resonant transitions, and occurs largely at the lower 
$1s2s (^3S_1,^1S_0)$ 
thresholds. The increase due to the higher 1s2pns series
limits $1s2p (^3P^o_{0,1,2},^1P^o_1)$ is much smaller. The wavelengths
corresponding to these threshold span the range 17.74 ($2^3S_1$) - 17.55
($2 ^1S_0, 2 ^3P^o_{0,1,2}$) - 17.42 $(2^1P^o_1)$ \ang.  
The O~VI resonant absorption cross section,
as $\lambda \rightarrow 17.42$ \ang, rises by a factor of 28 above the
background continuum cross section, 
approaching the $n = 2$ excitation thresholds of O~VII from below.
Of course this is also the energy region  corresponding to line opacity
from the O~VII $1s^2 \rightarrow 1snp$ transitions (Lee \etal include up
to $n = 5$ in their model shown in the bottom panel of Fig. 2).

 Also close to the \tov break are the lowest 
two resonances due to $2p^5 3s^2$ LMM complex of Fe~XVI. 
The first (stronger) LMM resonance lies at 17.38 \ang,
but the Fe~XVI bound-free continuum also has an appreciable value
at the \tov (the second LMM resoanance is at 17.10 \ang). 
Moreover, the opacity from the next Rydberg complex in 
Fe~XVI at $\sim$0.82 keV corresponds to the strong absorption feature
in the \chan \mcg6 spectrum (even higher $n$-complexes become weaker
roughly as $n^{-3}$).

\section{DISCUSSION}

 A few
salient points and implications of the present work are discussed.

 K- and L-shell resonant absorption from O and Fe ions, respectively, 
should be a significant contributor to the missing or unidentified 
X-ray opacity 
in the warm absorber models of \mcg6 by Lee \etal (2001) and Sako \etal (2002).
However the crucial question --- whether the additional resonance opacity 
is sufficient, or whether relativistic emission still needs to be invoked 
--- depends on the details of the model. In particular the
ionic column densities and O:Fe abundances need to explain, for exmaple,
the required optical depth of $\sim$0.7 at \tov. Nonetheless, it
appears that most if not all atomic features in the observed spectrum may be
theoretically modeled.

 It appears likely that the \ka resonances of all ionization stages of 
oxygen are present in the \chan \mcg6 spectrum.
However, this presents rather a challenge to model,
since the plasma source must necessarily imply a widely varying extended
environment and/or intervening matter along the line of sight to the
AGN. Whereas the O~VI resonance opacity spectrum bears a striking resemblance 
to the observed spectra at the \tov, complementing O~VII line opacity
(Fig. 2), 
the two ionization stages may differ by an order of magnitude
in temperature, depending on the preponderance of photoionization or
coronal equilibrium (or a hybrid) of the
(e~+~O~VII) $\rightleftharpoons$ O~VI system. We therefore
suggest an observational study of the O~VI UV \lmlm 1031,1038 \ang
doublet, cross-correlated with the O~VI X-ray resonant absorption
features described herein. 
The O~VI KLL doublet {\it absorption} resonances
at \lmlm  22.05 and 21.87 \ang (Pradhan 2000) lie
between the well known forbidden (`f'
or `z'), intercombination (`i' or `x,y'), and allowed (`r' or `w')
{\it emission} lines of O~VII due to
transitions $2(^3S_1, ^3P^o_{2,1}, ^1P^o_1)
 \longrightarrow 1(^1S_0)$ at \lmlm 22.101, 21.804, and
21.602 \ang respectively.
To augment theoretical studies,
 new self-consistent cross sections and rates O~VI/O~VII/O~VIII have been
computed (Nahar and Pradhan, in preparation) using the same
eigenfunction expansion for both photoionization and recombination, and
a unified (e~+~ion) recombination scheme including both radiative and
dielectronic recombination, in an ab initio manner using the
relativistic BPRM method
(e.g. Nahar and Pradhan 1995, Zhang \etal 1999).
The accurate new O~VII total and level-specific recombination rates and
A-values, up to $n$ =
10 fine structure levels, should generally help the study of recombination
lines, such as the O~VII `He$\beta, \gamma, \delta, \epsilon$' identified by
Sako \etal (2002) in the \xmm spectrum, and for which the model absorption
widths are `severely overpredicted'. 
Fine structure
collisional cross sections and rates for O~VII have also been
recently computed including
up to $n$ = 4 levels (Delahaye and Pradhan 2002).

 The shape of the oxygen gap in the observed flux at $\sim$
0.55 keV (22-23 \ang) is similar to 
the one in the $\sim$ 0.94 keV region where Lee
\etal (2001) have identified a line from He-like Ne~IX ($1s^2 -
1s2p$). Based on the present study, 
inner-shell excitation from {\it all} Ne-ions 
should contribute to a `neon gap' in this region. However, this
gap would overlay the much stronger Fe L-shell opacity at all energies higher
than the spectral break at \tov. Nonetheless, in general we might predict
similar gaps due to other elements, such as carbon at
$\sim$ 0.3 keV, nitrogen at $\sim$ 0.4 keV, neon at $\sim$ 0.9 keV, and
from inner-shell \ka resonances in iron in the 6.4 -
6.7 keV range (the \ka complex in He-like iron at 6.7 keV, and
Li-like dielectronic satellites at lower energies, 
are discussed by Oelgoetz and Pradhan 2001). These gaps or windows
due to \ka resonant excitation in all ionization stages of an element
would lie close together in the X-ray, and should 
be discernible (subject to the total opacity in that region).
 We note that the tentative identification by Lee \etal (2001) of the 
O~I resonance in \mcg6 appears to be 
slightly higher in energy than the present value, and 
previous experimental and theoretical
values at $\sim 23.5$ \ang discussed by
Paerels \etal (2001), who used the non-relativistic 
theoretical cross sections for O~I computed
by McLaughlin and Kirby (1998) to analyse interstellar X-ray absorption
(the theoretical model spectrum was shifted by
0.051 \ang to match the measured centroid wavelength of the O~I
resonance). Our O~II \ka resonance is about 0.08 \ang lower than inferred
from experimental data; however, new calculations are in progress.

 Line and resonance complexes are sometimes referred to as 
`unresolved transition
arrays' (UTA's, e.g. Sako \etal 2002). The present work demonstrates that it may be necessary to
highly resolve the UTA's, and precisely calculate the resonance profiles
and energies, to analyze the high resolution data from \chan and \xmm.  
Owing to the convergence of resonance series limits from below
threshold, the inner-shell ionization `edges' are {\it always} 
diffuse and exhibit
redward broadening. The total resonance averaged photoabsorption cross section 
is an analytic continuation of the cross section from just above threshold;
continuum lowering effects at resonance series limits should also be
manifiest under appropriate plasma conditions.

In Fig. 1 we have presented only the \ka photoabsorption 
cross sections from the ground level of O-ions.
Excited fine structure levels of the ground LS term may also be significantly
populated; resonances in
these excited level cross sections will also contribute to the
absorption spectrum. As mentioned, we are computing
comprehensive datasets for line $f$-values and photoionization cross
sections of O-ions and Fe~XVI, XVII, and XVIII, that should
provide a more complete
description of oxygen K-shell opacity for $\lambda > $17.5 \ang,
 and iron L-shell opacity for $\lambda < $17.5 \ang in the
region covered by both \chan and \xmm.

\section{CONCLUSION}

 Comparison of ab initio theoretical atomic parameters that 
determine the X-ray opacity 
of O and Fe with the high-resolution \chan and \xmm observations reveal 
some spectral features
heretofore unobserved in laboratory or astrophysical sources. We note
the following main conclusions:

\bl  Inner-shell \ka resonances from all O-ions O~I~-~OVI are
identified and associated with the window observed in the X-ray flux at
$\sim$ 22-23 \ang --- spectral signature of oxygen absorption from 
an `all-component' (mainly) optically thin plasma. Column densities of
all ionization stages may be obtained using the calculated
resonance oscillator strengths \fbar, and should provide stringent
constraints on parameters such as temperatures and ionization fractions
in photoioization models.

\bl Resonance series limits in O~VI, and LMM resonances in Fe~XVI,
should contribute significantly to the opacity at the spectral \tov
break in the 17.4 - 17.7 \ang range, possibly accounting for the missing 
O and Fe opacity to resolve the discrepancy between the
the relativistic redshift and broadening model, and the dusty warm
observer model. 

\bl In general, high  
precision resolution of resonances is essential for theoretical X-ray
models to match the high-resolution spectra of \chan and \xmm,
as astrophysical laboratories of fundamental atomic and
plasma physics.

\acknowledgments

 This work was partially supported by the NASA
Astrophysical Theory Program. The computational work was carried out on
the Cray SV1 at the Ohio Supercomputer Center in Columbus, Ohio.

\newpage

\def\r{\leftskip10pt \parindent-10pt \parskip0pt}
\def\apj{{ApJ}\ }
\def\apjs{{ApJS}\ }
\def\mn{{MNRAS}\ }
\def\aa{{A\&A}\ }
\def\aasup{{A\&A Suppl.}\ } 
\def\jpb{{Journal of Physics B}\ }
\def\pra{{Physical Review A}\ }
\def\prl{{Physical Review Letters}\ }
\def\adndt{{Atomic Data And Nuclear Data Tables}\ }
\def\cpc{{Comput. Phys. Commun.}\ }


\newpage

\begin{figure}
\centering
\psfig{figure=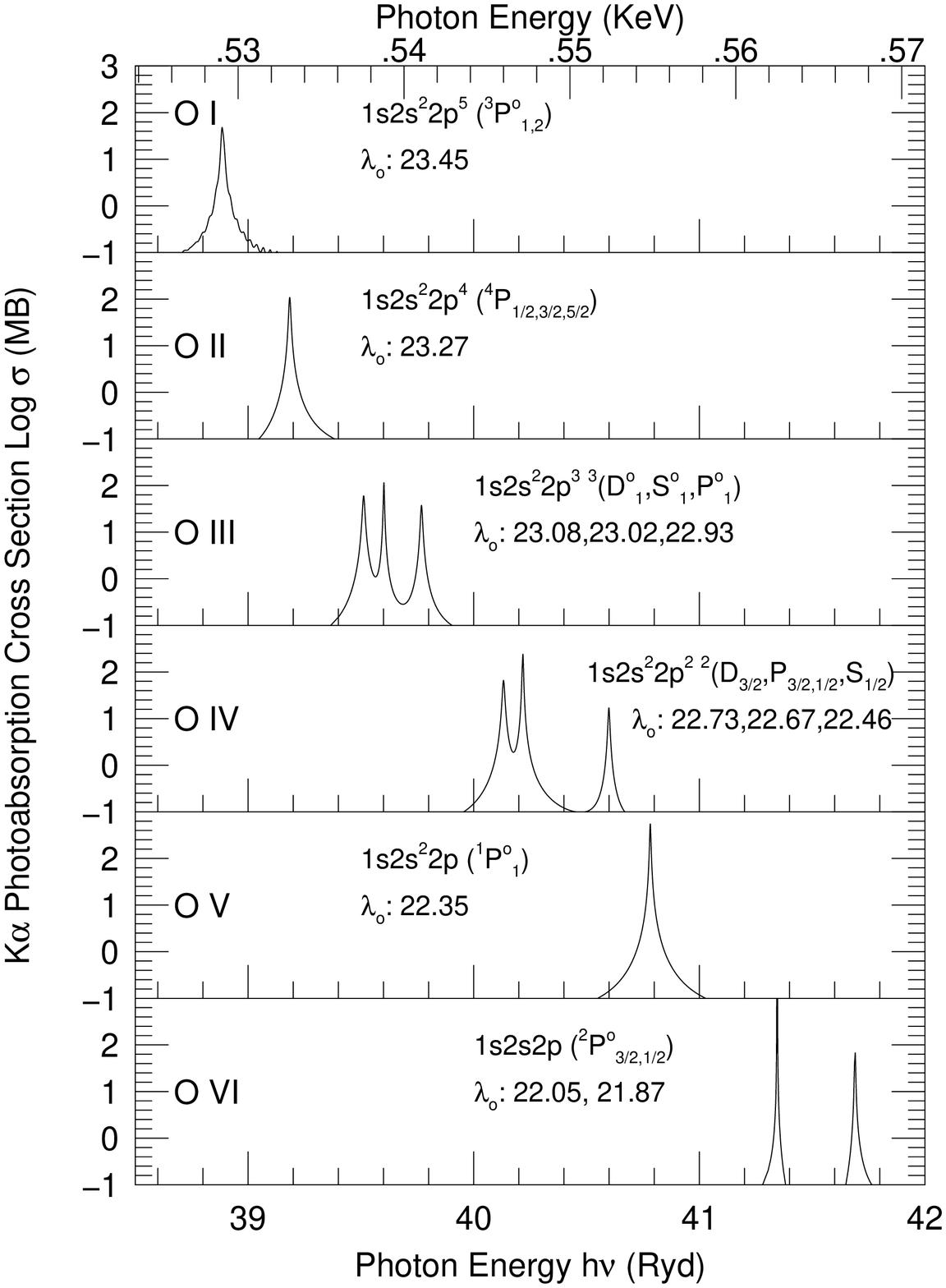,height=15.0cm,width=18.0cm}
\caption[f1.ps] {K$_{\alpha}$ photoabsorption cross sections of
resonance complexes in Oxygen ions O~I~-~O~VI.}
\end{figure}

\begin{figure}
\centering
\psfig{figure=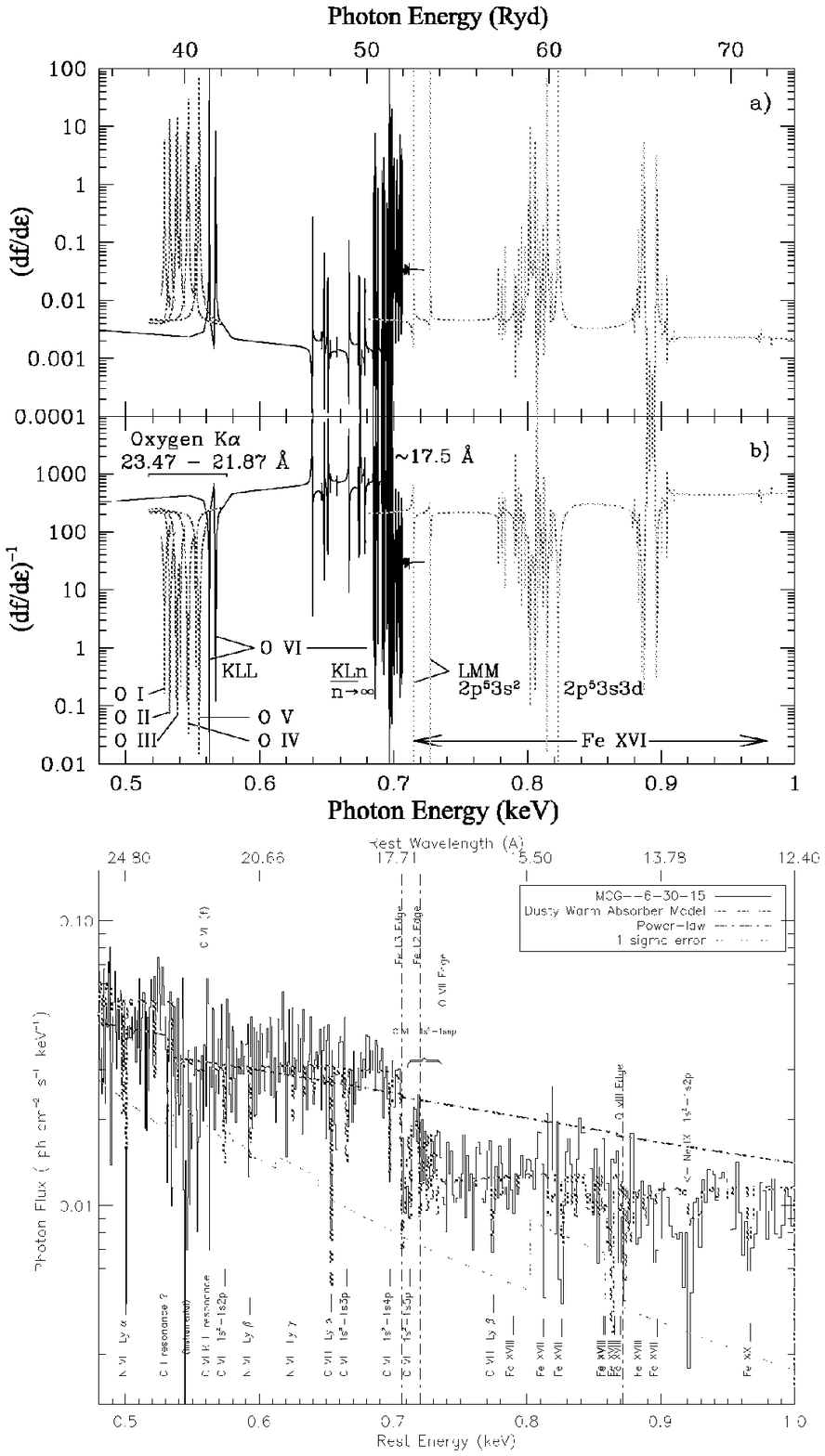,height=18.0cm,width=14.0cm}
\caption[f2.eps]{Differential oscillator strengths for
O-ions and Fe~XVI: (a) $df/d\epsilon$ --- O~I~-~V (red), O~VI (black),
Fe~XVI (green), (b) $ (df/d\epsilon)^{-1}$ compared with the observed
MCG6--6-30-15 spectrum by Lee \etal (2001) (bottom panel).}
\end{figure}

\begin{figure}
\centering
\psfig{figure=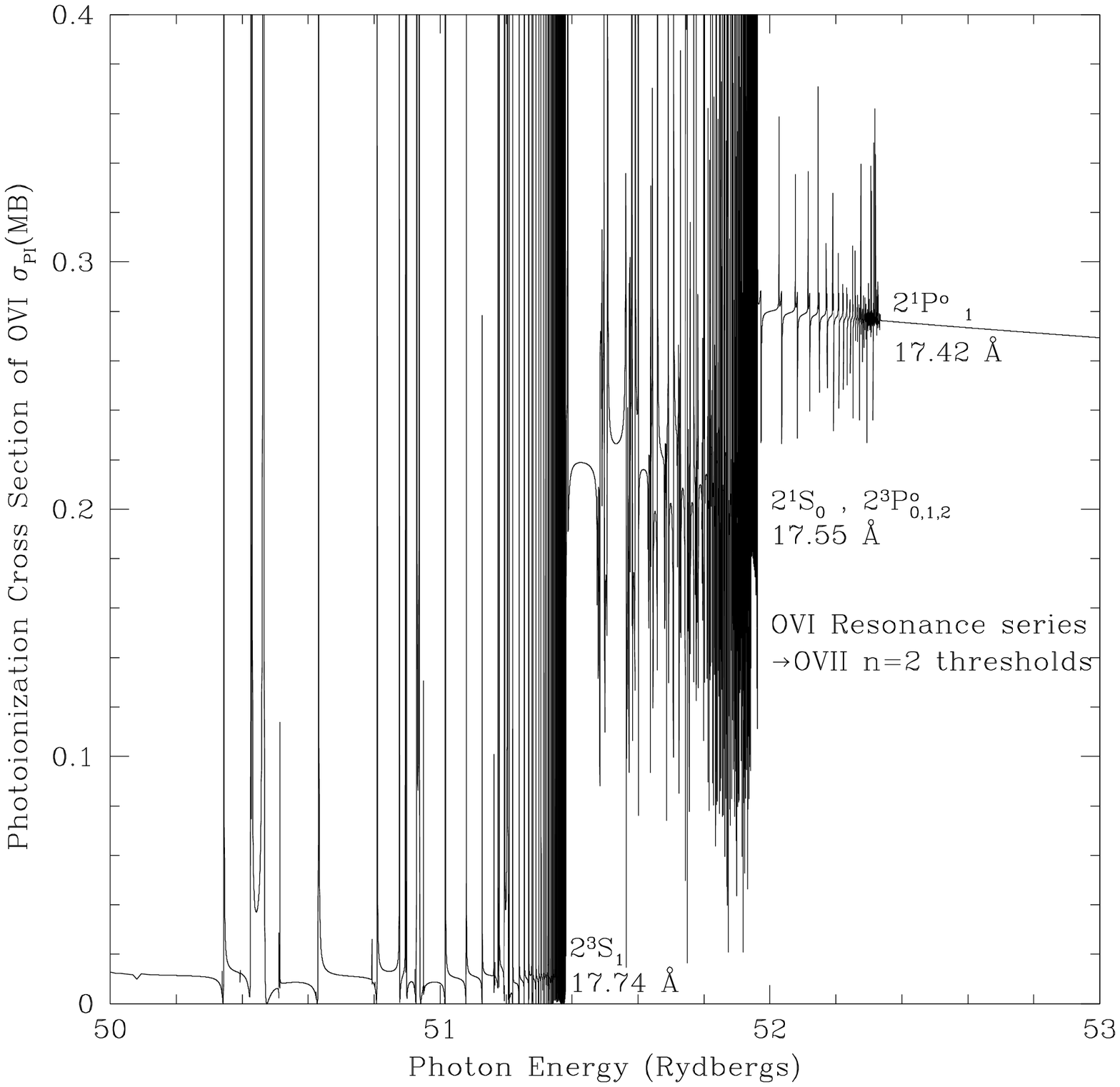,height=15.0cm,width=18.0cm}
\caption[f3.eps]{O~VI resonances at the $n$ = 2 O~VII series limits:
highly resolved photoionization cross sections in the $\sim$ 0.7 keV region.} 
\end{figure}

\end{document}